# Real-time stock analysis for blending recipes in industrial plants

Florin ZAMFIR
*Control Engineering, Computers & Electronics Department*
*Petroleum-Gas University of Ploiesti*
Ploiesti, Romania
florin.zamfir@upg-ploiesti.ro
https://orcid.org/0000-0003-2776-9234

Nicolae PARASCHIV
*Control Engineering, Computers & Electronics Department*
*Petroleum-Gas University of Ploiesti*
Ploiesti, Romania
nparaschiv@upg-ploiesti.ro
https://orcid.org/0000-0002-8923-0966

Emil PRICOP
*Control Engineering, Computers & Electronics Department*
*Petroleum-Gas University of Ploiesti*
Ploiesti, Romania
emil.pricop@upg-ploiesti.ro
https://orcid.org/0000-0002-4021-6549

*Abstract* — Many companies use Excel spreadsheets to keep stock records and to calculate process-specific data. These spreadsheets are often hard to understand and track. And if the user does not protect them, there is a risk that the user randomly changes or erase formulas. The paper focuses on the stocks of products used in a blending process with a known recipe. Developing an application that can bring this data in a centralized form and that can assist the operator in decide is a necessity. When a programmer implements an application that uses data from plants he needs to consider one fundamental aspect as reading real-time data from the process. The real-time stock analysis application takes into account all the above elements. The application is easy to use by an operator in the command room of installation because of the planning algorithms integrated into it. The algorithms proposed and implemented in this paper have well-defined goals: identifying the ingredients needed to achieve the blending process for required quantities, determine the quantities of the finished product that can be made with the existing ingredients and determine the optimum quantities of the finished product. The application implemented in C# intensively uses these algorithms and gives the user the ability to build the result step by step.

## I. INTRODUCTION

Planning is an essential feature of a refinery blending process based on the known recipe (specifications). The infrastructure limits the number of unique recipes a refinery can produce. More recipes mean an increasing complexity of the process infrastructure. With such a complex system, the stock analyzes can be a powerful instrument of production planning. Many times, the delays from a production process can come from subjective reasons, such as lack of communication between departments [8].

A blending process must consider the following [6]:

- stocks of raw materials must be sufficient to carry out the blending process;
- the existence of automation systems for transporting the components from vessels to the mixing process is fundamental;
- blending recipes must be correctly defined;
- changing a specific recipe for a particular blended product can lead to another blended product;
- components must be chosen in such a way as to obtain maximum profit.

A control system for the blending process must contain the following elements: mixing chamber, pipelines for the incoming liquids transport, pipelines for the mixed products transport, sensors, controller connected to the process elements [2].

With regard to the planning of blending processes, three types of activities can be distinguished within a company: process control, short-term planning, and long-range planning. Long-range planning refers to a company objective defined through permanently problems solving and reaching their targets. Short-term planning is about getting immediate results, results that can lead to the ultimate goal of planning. Process control has a major impact on the other two types of activities, meaning that the blending process can influence decisions on short or long term planning. At the same time, a decision on reaching certain deadlines or goals may affect the blending process [6][7].

For blending systems design (fig. 1) it is important to consider the development of a subsystem that can provide process information to the operator. Fig. 1 illustrates a more simplified version of the blending process model from [4].

The data from the process are simple process parameters (tank level, pipe flow, pressure, etc.) or complex information about blending planning. For example, the estimation of the ingredients needed for the achievement of a certain quantity of the blended product based on linear programming can be the first stage of such a system. The second stage comprises checking the stock of ingredients and identifying the optimal quantity of blended products that will derive from them.

Fig. 1. Blending plant with multiple recipes

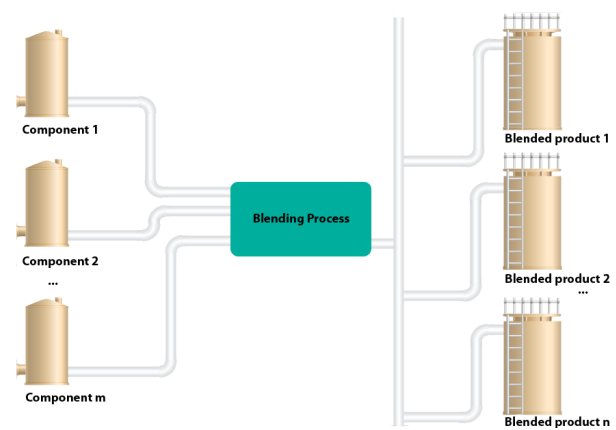

The proposed solution requires the utilization of a wireless sensor network connected through LoRa technology, capable of data acquisition from the process. All the sensor from the network needs to connect to a Lora gateway that can facilitate the data transmission to the server. A series of blending algorithms process the data transmitted, and the operator receives the results.

II. PLANNING ALGORITHMS FOR THE BLENDING PROCESS

The authors proposed three algorithms that can aid the planning application to identify the best solution based on process status.

The first algorithm (A) proposed and implemented can identify the number of components required to produce a certain quantity of blended product. The second algorithm (B) can calculate the maximum quantities of a certain blended product that can be made using available components. The third algorithm (C) can determine the optimal quantities of blended products that can be made with all recipes in use based also on the available components.

*A. Algorithm for components identification required to produce a certain quantity of blended product.*

The recipes of the blended products are like a matrix where rows describe the number of blended products used in the mixture of m components (m columns). The values of the matrix A from (1) represents the weight of components in the blended product (the sum of each row must be 1).

$$A = \begin{bmatrix} a_{11} & a_{12} & \cdots & a_{1m} \\ a_{21} & a_{22} & \cdots & a_{2m} \\ \vdots & \vdots & \ddots & \vdots \\ a_{n1} & a_{n2} & \cdots & a_{nm} \end{bmatrix} \quad (1)$$

The required quantities of blended products (QB) can be described as well as a matrix with n rows and one column as in (2).

$$P = \begin{bmatrix} p_1 \\ p_2 \\ \vdots \\ p_n \end{bmatrix} \quad (2)$$

Equation (3) express the requirement of quantities of components (QC) used for blending a given quantity of blended products.

$$C = \begin{bmatrix} c_{11} & c_{12} & \cdots & c_{1m} \\ c_{21} & c_{22} & \cdots & c_{2m} \\ \vdots & \vdots & \ddots & \vdots \\ c_{n1} & c_{n2} & \cdots & c_{nm} \end{bmatrix} \quad (3)$$

In order to obtain the matrix C, the algorithm needs to multiply the recipe matrix A[n x m] and the required quantities of blended product from matrix P[n x 1]. But this operation cannot be done. The solution is to transform the matrix P[nx1] into a matrix B[n x n].

$$\begin{bmatrix} b_{11} & b_{12} & \cdots & b_{1n} \\ b_{21} & b_{22} & \cdots & b_{2n} \\ \vdots & \vdots & \ddots & \vdots \\ b_{n1} & b_{n2} & \cdots & b_{nn} \end{bmatrix} \cdot \begin{bmatrix} a_{11} & a_{12} & \cdots & a_{1m} \\ a_{21} & a_{22} & \cdots & a_{2m} \\ \vdots & \vdots & \ddots & \vdots \\ a_{n1} & a_{n2} & \cdots & a_{nm} \end{bmatrix} = \begin{bmatrix} c_{11} & c_{12} & \cdots & c_{1m} \\ c_{21} & c_{22} & \cdots & c_{2m} \\ \vdots & \vdots & \ddots & \vdots \\ c_{n1} & c_{n2} & \cdots & c_{nm} \end{bmatrix} \quad (4)$$

The coefficients of matrix B from (5) can be obtained after a series of matrix algebra operations. These coefficients represent the rows from matrix P[n x 1] placed in a diagonal order into a new matrix named $P_1$ [n x n].

$$\begin{bmatrix} p_1 & 0 & \cdots & 0 \\ 0 & p_2 & \cdots & 0 \\ \vdots & \vdots & \ddots & \vdots \\ 0 & 0 & \cdots & p_n \end{bmatrix} \cdot \begin{bmatrix} a_{11} & a_{12} & \cdots & a_{1m} \\ a_{21} & a_{22} & \cdots & a_{2m} \\ \vdots & \vdots & \ddots & \vdots \\ a_{n1} & a_{n2} & \cdots & a_{nm} \end{bmatrix} = \begin{bmatrix} c_{11} & c_{12} & \cdots & c_{1m} \\ c_{21} & c_{22} & \cdots & c_{2m} \\ \vdots & \vdots & \ddots & \vdots \\ c_{n1} & c_{n2} & \cdots & c_{nm} \end{bmatrix} \quad (5)$$

The algorithm can be described now based on mathematical implementation. Fig.2 shows pseudocode implementation of matrix C[n x m] calculation. The function used has as inputs the matrices A and P and returns the resulting matrix C as output.

Fig. 2. Pseudocode algorithm implementation for the identification of the components

```
DEFINE recipes matrix A and required blended
    products matrix P;
FUNCTION component_id
    Pass In: matrix A, matrix P
        COMPUTE matrix P1 using the described
        relations;
        COMPUTE matrix C based on (5);
    Pass Out: matrix C
ENDFUNCTION
CALL: C =component_id(A,P);
PRINT A,P and C matrix.
```

The first set of instructions in the component_id() function consists of processing the matrix P[n x 1] and writing it as a matrix P1[n x n]. The second set computes the matrix C[n x m] using the relations in (5).

*B. Algorithm for maximum quantities calculation of a certain blended product*

For this algorithm is necessary to identify a matrix of blended products $P_2$ [n x 1] where each row is equal with the maximum quantity of blended product that can be made with the recipes matrix A[n x m] and the available components (QC) matrix $C_2$ [1 x m].

This algorithm will return in matrix $P_2$ all the possible maximum QP that can be obtained from the available QC.

This time, the relation between the three matrices cannot be written with rules of matrix algebra. In these conditions, it is required to implement the function from (6) that can resolve the problem. It is assumed that the P2 matrix is a function of A and C2.

$$P_2 = f(A, C_2) \quad (6)$$

Relation (6) can be written in a matrix form:

$$\begin{bmatrix} p_1 \\ p_2 \\ \vdots \\ p_n \end{bmatrix} = \begin{bmatrix} a_{11} & a_{12} & \cdots & a_{1m} \\ a_{21} & a_{22} & \cdots & a_{2m} \\ \vdots & \vdots & \ddots & \vdots \\ a_{n1} & a_{n2} & \cdots & a_{nm} \end{bmatrix} \diamond \begin{bmatrix} c_1 & c_2 & \cdots & c_m \end{bmatrix} \quad (7)$$

The symbol " $\diamond$ " represents the function *max_quantities()* implemented in the pseudocode from *Fig.*3. The value *minimum_bp* from the same fig. 3. represents the value of the current blended products that is defined as an integer-part the determined value of the report between the available component and the weight of the recipe from the current iteration. The function *max_quantities()* has a set of rules that stores in each row of matrix P2 the smallest value of variable *minimum_bp*. The minimum_bp variable is calculated by reference to all columns of the recipe matrix.

Fig. 3. Algorithm for maximum quantities determination

```
DEFINE recipes matrix A and available components
        matrix C2
FUNCTION max_quantities
  Pass In: matrix A, matrix C2
    SET line = 1 ;
    FOR each line in A
        SET column = 1 and prod = 9999;
        FOR each column in A
            SET  weight = A(line,column);
            SET available_comp = C2(column);
            SET minim_bp = [available_comp/weight ];
            IF prod > minim_bp AND weight !=0
                SET prod = minim_bp;
                INCREMENT Column ;
            SET P2(line,1) = prod;
            INCREMENT line;
        END FOR
    END FOR
  Pass Out: matrix P2
ENDFUNCTION
    CALL: P2 = max_quantities(A,C2);
PRINT A,P2 and C2 matrix.
```

According to the function, after identifying the minimum amount that can be achieved with all the ingredients, the current row of the P2 matrix will be updated with that value.

### C. Algorithm for calculating the optimal amount of blended products made with the available ingredients

The first algorithms (A and B) are useful for the blending process but are not enough. Often, a simple analysis of the required components used to make a particular blend cannot provide information for the entire process. Also, the results of the second algorithm are not enough for the operator to make a decision. This is why a new algorithm is needed to identify, according to the user's requirements, the maximum quantities of blended products that can be made. In order to do this, a function that returns information about the current state of the components must be defined.

Fig. 4. Algorithm for optimal quantities of blended products determination

```
FUNCTION component_state
  Pass In: matrix C, matrix C2, int ColNoA
        SET C2Used = sum of each column from C;
        SET C2Temp = C2;
        SET C2 = C2 – C2Used;
        SET negative = 0;
        FOR i = 1 to  ColNoA
           C2Required (i)=0;
           IF C2(i)<0
              negative =1;
              C2Required (i) = C2Used(i) – C2Temp(i);
        END FOR
  Pass Out: matrix C2Used, matrix C2, matrix
  C2Required, bool negative
ENDFUNCTION
```

The *component_state()* function has as input data the matrix C calculated by the *component_id()* function and the matrix C2 containing the information about the ingredients inventory. The function returns several status parameters of the blending process.

The matrix C2Used stores information about the components spent to make the required blend. It also provides an updated version of the matrix C2.

Fig. 5 shows in pseudocode the optimization algorithm that uses all the functions described above. The first step is to calculate matrix C. The second step involves identifying the current state of the blending process.

Fig. 5. Algorithm for optimal quantities of blended products determination

```
DEFINE recipes matrix A, available components matrix
        C2 and required blended products matrix P;
COMPUTE RowNoA, ColNoA, RowNoC2, ColNoC2
CALL: C = component_id(A,P);
CALL: component_state(C,C2, ColNoA)
IF (negative ==1)
   PRINT ("Required blended products cannot be made" )
   PRINT matrix C2Required;
ELSE
   CALL: P2 = max_quantities(A,C2);
   P2Temp =P2;
   FOR i = 1 to  RowNoA
      FOR j = 1 to  RowNoA
          IF j not equal i
             P2(j) = 0;
      END FOR
      FOR k =1 to   ColNoA
          IF P(i,k) not equal 0
             CALL: C = component_id(A,P2)
             CALL: component_state(C,C2, ColNoA)
             CALL: P3 =max_quantities(A, C2Required);
             P2 = P2+P3;
      END FOR
   END FOR
Ptotal = P + P2;
Print Ptotal matrix
```

This algorithm must also contain a verification mechanism. The function component_state() from fig. 4 must also compute the number of ingredients missing to provide the blended products required by the user. This event is brought to the user attention via the flag "negative".

### III. THE PLANNING APPLICATION FOR THE BLENDING PROCESS

The program that implements the algorithms A, B and C is illustrated in fig. 6. The application uses the C # programming language. To illustrate the method of functioning, it is considered a recipe matrix consisting of two mixtures. The first blending product uses all nine components and the second one uses the first six components. At the click of the "*Upload Recipes*" button, information about blending recipes from a CSV file is sent to A[2 x 9] matrix. The first column of each row in the CSV file contains the name of the blended product and the remaining columns provide the required amount of components to achieve a ton of that product.

The "*Download Available Quantities of Database from Database*" button brings the latest version of the information collected from the sensors into C2[1 x 9] matrix (the number of semi-finished products stored in the tanks).

Fig. 6. Real-Time Stock Analysis main interface

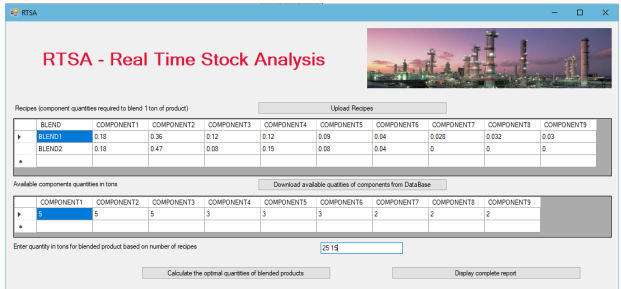

The application is built so that the operator can easily intervene both to modify the recipe and to modify the need for the semi-finished product. Operator intervention may involve changing recipe values or even adding new recipes, as can be seen in fig.7.

Fig. 7. Modifying values directly in RTSA application

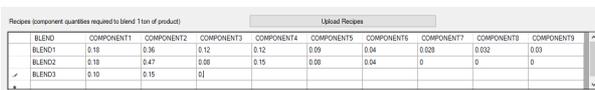

For the data entered in the interface from fig.6, the production of 25 tons of BLEND1 and 15 tons of BLEND2 is desired. The message from *Fig.*9 will be displayed when pressing the button "*Calculate the optimal quantities of blended products*".

Fig. 8. Real-Time Stock Analysis quantity overflow error

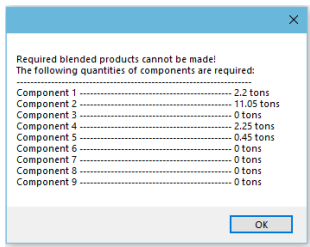

This message appears because the quantities required by blended products exceed the quantities available in the tanks.

If the operator chooses to produce 5 tons of BLEND 1 and 3 tons of BLEND2, a new window will open as in *Fig.*10 that will guide the user to the next confirmation steps.

At the top of the window are displayed the quantities of the components needed to produce what the operator intends. In the bottom left section, the quantities available in the tanks after blending are displayed. The central section of the window shows the operator the possible work variants with the remaining ingredients.

In the section on the bottom right, the operator can enter the numerical value for the version. For example, if he chooses option 1, the application considers that the operator wants to produce another 4 tons of BLEND1 and will calculate if other quantities can still be produced with the remaining ingredients.

Fig. 9. Real-Time Stock Analysis reports interface

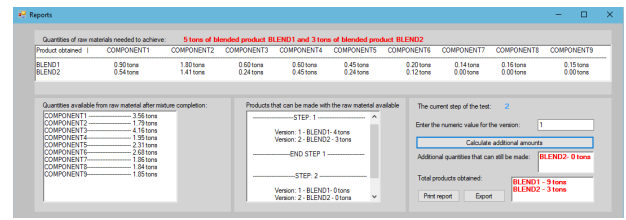

When pressing the "*Calculate additional amounts*" button, the current step increments automatically and the next possible choices appear in the center section. But in this case for step 2 both versions returned value zero so the iterations will not continue.

It shows that nothing more can be done. In the last part, the operator can see the total blended products that can be obtained with the ingredients available in the tanks.

The interface from fig.9 also contains another two important buttons. The "*Print report*" button prints the data in an easy-to-follow form and the Export button allows the operator to generate an editable file containing the chosen solution. In the main application interface of *Fig.*6, there is also a button that allows export of data for all possible variants. This allows the pursuit of all solutions and the choice of the best option.

IV. RESULTS

Next, a test with a recipe matrix containing three mixing products will be performed as can be seen in Fig.10. The quantities of ingredients remain constant and the operator chooses to produce the following quantities: 1 ton of BLEND1, 2 tons of BLEND2 and 4 tons of BLEND3.

Fig. 10. RTSA application test with three blending recipes

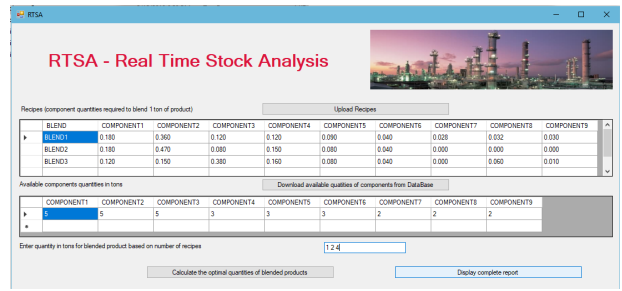

By pressing the "*Display complete report*" button, the user will be able to view a file that includes all the steps with all the possible combinations for the three blended products.

Fig.11 and table I summarize the results obtained. Each iteration continues until the P3 matrix becomes zero. All leaves of the tree will have the value of [0 0 0]. The red-colored values in the tree nodes confirm that no blended products can be made. At that time, it is possible to determine the total quantities of blended products taking into account the imposed conditions. The initial quantities of components for the root of the tree from fig.12 are determined after the required amount has been deducted to achieve the values required by the operator. Matrix C2 is updated based on the chosen branch, and based on these, the values for matrix P3 are determined.

Fig. 11. Hierarchical calculation of component requirements

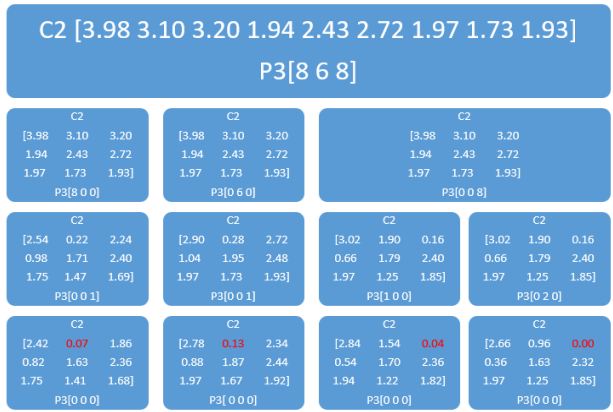

Table I highlights the optimization of the blending process by updating the quantities of blend products obtained at each branch of the tree from fig.11.

TABLE I. VALUES FOR P2 AND P3 MATRIX FOR ALL ITERATIONS

| step | sub step | P3 matrix | P2 Possible blended products | | |
|---|---|---|---|---|---|
| | | | BLEND1 (tons) | BLEND2 (tons) | BLEND3 (tons) |
| 0 | 0 | [1 2 4] | 1 | 2 | 4 |
| 0 | Possible choises: | | 8 | 6 | 8 |
| 1 | 1 | [8 0 0] | 9 | 2 | 4 |
| 1 | 2 | [0 0 1] | 9 | 2 | 5 |
| 1 | 3 | [0 0 0] | **9** | **2** | **5** |
| 2 | 1 | [0 6 0] | 1 | 8 | 4 |
| 2 | 2 | [0 0 1] | 1 | 8 | 5 |
| 2 | 3 | [0 0 0] | **1** | **8** | **5** |
| 3 | 1 | [0 0 8] | 1 | 2 | 12 |
| 3 | Possible choises: | | 1 | 2 | 0 |
| 3 | 1.1 | [1 0 0] | 2 | 2 | 12 |
| 3 | 1.2 | [0 0 0] | **2** | **2** | **12** |
| 3 | 2.1 | [0 2 0] | 1 | 4 | 12 |
| 3 | 2.2 | [0 0 0] | **1** | **4** | **12** |

Four variants are presented to the operator. Using available stock of material, the plant can run with one of four suggestions:

- 9 tons of BLEND1, 2 tons of BLEND2 and 5 tons of BLEND3;
- 1 ton of BLEND1, 8 tons of BLEND2 and 5 tons of BLEND3;
- 2 tons of BLEND1, 2 tons of BLEND2 and 12 tons of BLEND3;
- 1 ton of BLEND1, 4 tons of BLEND2 and 12 tons of BLEND3.

The complete report file also contains detailed information for each iteration, such as state variables, ingredient matrix values, etc.

## V. CONCLUSIONS

Implementing this solution in a blending plant can be beneficial for production optimization. Adding a solution that can provide helpful aid to the operator can increase productivity.

The authors proposed three algorithms for the blending of the quantities of components with the finished products for the blending process.

As seen in sections III and IV, the application is scalable. But, in order for the algorithms to function correctly, the matrix structure of the input data must be retained This rule does not impose any restrictions because any recipe can be written in the above-mentioned form. These algorithms can be used for the calculation of gasoline and diesel recipes in a refinery. Another use would be in the food industry for recipes for mixing soft drinks or even alcoholic beverages.

A possible improvement of the application can be achieved by including functions that can associate cost for ingredients and sales price for finished products. In this way, economic predictions can be included, which will have a very high weight in the final decision taken by the operator.